# The Production of $^{44}$Ti and $^{60}$Co in Supernova

F. X. Timmes[1,2,3,4], S. E. Woosley[2,3], D. H. Hartmann[4], & R. D. Hoffman[2]

[1] Laboratory for Astrophysics and Space Research
University of Chicago, Chicago, IL 60637

[2] University of California Observatories / Lick Observatory
Board of Studies in Astronomy and Astrophysics
University of California, Santa Cruz, Santa Cruz, CA 95064

[3] General Studies Division, Lawrence Livermore National Laboratory
Livermore, CA 94550

[4] Department of Physics and Astronomy
Clemson University, Clemson, SC 29634

email: fxt@burn.uchicago.edu

ABSTRACT

The production of the radioactive isotopes $^{44}$Ti and $^{60}$Co in all types of supernovae is examined and compared to observational constraints including Galactic $\gamma$–ray surveys, measurements of the diffuse 511 keV radiation, $\gamma$–ray observations of Cas A, the late time light curve of SN 1987A, and isotopic anomalies found in silicon carbide grains in meteorites. The (revised) line flux from $^{44}$Ti decay in the Cas A supernova remnant reported by COMPTEL on the Compton Gamma-Ray Observatory is near the upper bound expected from our models. The necessary concurrent ejection of $^{56}$Ni would also imply that Cas A was a brighter supernova than previously thought unless extinction in the intervening matter was very large. Thus, if confirmed, the reported amount of $^{44}$Ti in Cas A provides very interesting constraints on both the supernova environment and its mechanism. The abundances of $^{44}$Ti and $^{60}$Co ejected by Type II supernovae are such that gamma-radiation from $^{44}$Ti decay SN 1987A could be detected by a future generation of gamma-ray telescopes and that the decay of $^{60}$Co might provide an interesting contribution to the late time light curve of SN 1987A and other Type II supernovae. To produce the solar $^{44}$Ca abundance and satisfy all the observational constraints, nature may prefer at least the occasional explosion of sub-Chandrasekhar mass white dwarfs as Type Ia supernovae. Depending on the escape fraction of positrons due to $^{56}$Co made in all kinds of Type Ia supernovae, a significant fraction of the steady state diffuse 511 keV emission may arise from the annihilation of positrons produced during the decay of $^{44}$Ti to $^{44}$Ca. The Ca and Ti isotopic anomalies in pre-solar grains confirm the production of $^{44}$Ti in supernovae and that extensive mixing between zones has occurred, but a quantitative model for this mixing is presently lacking.

*Subject Headings*: gamma rays: theory – nuclear reactions, nucleosynthesis, abundances – supernovae: general

1. Introduction

One byproduct of Galactic chemical evolution is the production of traces of short to medium lived radioactivities having lifetimes ranging from several years to several million years. Detecting the $\gamma$–radiation from the decay of these species, either in individual supernova remnants or through the glow of accumulated nucleosynthetic products in the interstellar medium provides a direct way to study the nature of supernovae and to calibrate nucleosynthesis theory (e.g., Clayton, Colgate, & Fishman 1969; Clayton 1974, Ramaty & Lingenfelter, 1977). In a previous paper we analyzed the production of two long lived isotopes $^{26}$Al and $^{60}$Fe (Timmes et al. 1995b). Here we concentrate on two shorter lived isotopes, $^{44}$Ti (half–life of 58 ± 10 yr; Meißner et al. 1995) and $^{60}$Co (half–life of 5.27 years; Tuli 1990). In addition to their gamma-radiation, these species have additional observational consequences for the late time (one year to several centuries) light curves of supernovae and for isotopic patterns measured in presolar grains. Moreover, since their decay lifetimes are comparable to or shorter than the average time interval between supernovae, their abundances can serve as important tracers of the most recent supernovae in our Galaxy. In this paper we summarize the production levels of these two isotopes in all types of supernovae and discuss the observational implications.

Our starting point is a grid of massive star models recently followed through their presupernova evolution by Weaver & Woosley (1996) and exploded, in a parametric way, by Woosley & Weaver (1995). Each star was exploded using a piston to give a specified final kinetic energy of the ejecta at infinity (typically $1.2 \times 10^{51}$ erg). In both the presupernova and exploded models the nucleosynthesis of 200 isotopes was determined. These give ejecta whose composition, when integrated into a Galactic chemical evolution model, compares favorably with what is observed in the Sun and other nearby stars (Timmes, Woosley & Weaver 1995a). We address the effects of mass loss by also including the stellar models of Woosley, Langer, & Weaver (1993, 1995), which explicitly followed the mass loss that might be appropriate for either single or mass exchanging binary stars evolved to supernovae. The mass of $^{44}$Ti and $^{60}$Co ejected is only affected if these stars lose not only all of their hydrogen envelopes, but end up with significantly smaller helium cores. Such a reduction in helium core mass seems necessary to produce the light curves of Type Ib supernovae (Ensman & Woosley 1988).

We also examine the synthesis of $^{44}$Ti and $^{60}$Co by various models for Type Ia supernova. Thermonuclear explosions of accreting carbon–oxygen white dwarfs that have approached the Chandrasekhar limit eject a mass of $^{44}$Ti that is typically, though not always, substantially smaller than the yields from Type II/Ib supernova. Models of helium detonations on sub–Chandrasekhar mass carbon–oxygen white dwarfs can synthesize large amounts of $^{44}$Ti (Woosley, Taam & Weaver 1986; Livne & Glasner 1991; Woosley & Weaver 1994; Livne & Arnett 1995). If explosion of these types of Type Ia supernova are not rare, they may be the main origin of $^{44}$Ca in the Galaxy.

The main result of this paper is to show consistency between the masses of $^{44}$Ti and $^{60}$Co ejected in current supernova models and several independent sets of observations - namely Galactic $\gamma$–ray surveys, measurements of the diffuse 511 keV radiation, $\gamma$–ray observations of Cas A, the late time light curve of SN 1987A, and the isotopic anomalies found in silicon carbide grains in meteorites.

## 2. The Production of $^{44}$Ti and $^{60}$Co in Supernovae

In massive stars, $^{44}$Ca is produced chiefly, though not exclusively, as the radioactive progenitor $^{44}$Ti made in the "$\alpha$–rich freeze–out". This process occurs when material initially in nuclear statistical equilibrium (NSE) at a relatively low density is cooled rapidly enough that the free $\alpha$–particles present in the NSE ensemble do not have time to merge back into the iron group by the relatively inefficient triple–$\alpha$ reaction. Thus, the distribution of nuclei cools down in the presence of an anomalously large concentration of $\alpha$–particles (Woosley, Arnett & Clayton 1973; Thielemann, Hashimoto & Nomoto 1990; Woosley & Hoffman 1991). The isotope $^{60}$Co is made by neutron capture in the neon and oxygen layers, so its synthesis is less sensitive to uncertainties in the explosion model.

Yields of $^{44}$Ti and $^{60}$Co from the (solar metallicity) Type II supernova models of Woosley & Weaver (1995) are shown in Fig. 1a, and from the Type Ib models of Woosley, Langer & Weaver (1995) in Fig. 1b. Production in Type Ib supernovae is more uniform because the models converge to a common presupernova mass in the narrow range 2.26 – 3.55 $M_\odot$. All of the models shown in Figs. 1a and 1b, whose ejecta all have nearly $10^{51}$ erg of kinetic energy at infinity, predict $^{44}$Ti yield between $10^{-5}$ and $1.5 \times 10^{-4}$ $M_\odot$, with typical values $\sim 3 \times 10^{-5}$ $M_\odot$ for the Type II models and $\sim 6 \times 10^{-5}$ $M_\odot$ for the Type Ib models. Note that the final mass of the helium core is a more appropriate parameterization than either the main–sequence or initial helium core mass of the star. The supernova models also predict a yield of $^{60}$Co between $10^{-6}$ to $10^{-4}$ $M_\odot$, with typical values being $\sim 3 \times 10^{-5}$ $M_\odot$ for the Type II models and $\sim 2 \times 10^{-6}$ $M_\odot$ for the Type Ib models.

Producing more than $10^{-4}$ $M_\odot$ of $^{44}$Ti is difficult in core collapse models, but making much less is easy. Indeed, ejection of *any* $^{44}$Ti is sensitive to how much mass falls back onto the remnant (Woosley & Weaver 1995). The mass of $^{44}$Ti external to the piston and the mass ejected is shown in Table 1 as a function of the main–sequence progenitor mass. The difference is the portion that falls back onto the neutron star. Entries in the table are for the models where the ejecta's kinetic energy is a nearly constant $10^{51}$ erg. The masses external to the piston are probably realistic upper bounds. In all cases, the mass external to the piston is greater than or equal to the ejected mass, with equality holding only for the least massive stars (11 and 12 $M_\odot$ in Table 1). To achieve a nearly constant kinetic energy of the ejecta, the explosion energy in the $M \geq 25$ $M_\odot$ presupernova models must be steadily increased in order to overcome the increased binding energy of the mantle. In the most massive stars (35 and 40 $M_\odot$ in Table 1) the binding energy and density profile causes nearly all the $^{44}$Ti produced to fall back onto the compact remnant (see Woosley & Weaver 1995 for a discussion of this fall back). Unless the explosion mechanism, for unknown reasons, provides a much larger characteristic energy in more massive stars, it appears likely that stars larger than about 30 $M_\odot$ will have dramatically reduced $^{44}$Ti yields and leave massive remnants ($M \geq 10$ $M_\odot$) which become black holes. If, however, the explosion is energetic enough (perhaps overly energetic) more $^{44}$Ti is ejected. Clearly, other than the upper bounds of Table 1, the $^{44}$Ti synthesis in models of very massive stars is quite uncertain.

In the solar metallicity 25 $M_\odot$ model, the isotope $^{60}$Co, like $^{60}$Fe, is made chiefly in the oxygen and neon shells in the last stages of presupernova evolution (see Fig. 2a) and to a lesser extent in a thin layer at the base of the He-burning shell. The production of $^{60}$Fe by neutron capture in the explosive helium burning shell is larger than the production of $^{60}$Co (made also by neutron capture, see Fig. 2b), since the $^{56,58}$Fe seed nuclei are more abundant than the $^{59}$Co seed nuclei. The difference between production by neutron exposure in the helium shell and in the convective neon–oxygen shell may be shown by examining the integrated totals. Exterior to 6.53 $M_\odot$ (the bottom of the He shell) there is $1.9 \times 10^{-7}$ $M_\odot$ of $^{60}$Fe and $3.4 \times 10^{-6}$ $M_\odot$ of $^{60}$Co in the 25 $M_\odot$ presupernova model and $1.2 \times 10^{-5}$ $M_\odot$ of $^{60}$Fe and $4.5 \times 10^{-6}$ $M_\odot$ of $^{60}$Co after the explosion. Between 2.07 $M_\odot$ (the future mass cut) and 6.53 $M_\odot$ the presupernova has $7.8 \times 10^{-6}$ $M_\odot$ of $^{60}$Fe and $4.7 \times 10^{-5}$ $M_\odot$ of $^{60}$Co, while after the explosion there is $9.4 \times 10^{-6}$ $M_\odot$ of $^{60}$Fe and $4.9 \times 10^{-5}$ $M_\odot$ of $^{60}$Co. Thus, the total ejected mass of $^{60}$Fe is $2.1 \times 10^{-5}$ $M_\odot$ and $5.4 \times 10^{-5}$ $M_\odot$ of $^{60}$Co as given in the tables of Woosley & Weaver (1995). In addition to its sensitivity to explosion characteristics and the strength of the s-process, the amount of $^{60}$Co made here might be altered by a more careful treatment of convection. Bazan & Arnett (1994) find that plume structures dominate the two dimensional velocity field of convective oxygen shell burning and that significant local hot spots of nuclear burning exist due to mixing beyond the boundaries defined by mixing-length theory. The composition inhomogeneities and local burning are likely to change the quantitative yields of several isotopes from a single star as compared to one–dimensional calculations. Yields from a well defined progenitor mass may even be mildly chaotic.

Since we are dealing with neutron rich nuclei near the boundary of the network employed by Woosley & Weaver (1995), it was desirable to check the results using a larger set of reactions. The solar metallicity 15 and 25 $M_\odot$ stars were evolved with a larger nuclear reaction network (500 isotopes, up to and including the isotopes of ruthenium). Also included was a temperature dependent beta decay rate for $^{60}$Co which effectively decreases its synthesis during helium burning. After the entire evolution (presupernova + explosion) the mass of $^{44}$Ti ejected by the 15 M$_\odot$ star decreased 4.9% and the mass of $^{60}$Co ejected decreased by 25%. For the 25 M$_\odot$ model, $^{44}$Ti increased by 25% while $^{60}$Co decreased by -3.1%. We conclude that the error in the abundances of these isotopes due to the smaller size of network used by Woosley & Weaver (1995) is negligible.

Deflagration models for Type Ia supernovae in which carbon ignites at or near the center of a carbon-oxygen white dwarf of 1.39 M$_\odot$ produce $1.8 \times 10^{-5}$ M$_\odot$ of $^{44}$Ti and $2 \times 10^{-8}$ M$_\odot$ of $^{60}$Co (Nomoto, Thielemann & Yokoi 1984; Thielemann, Hashimoto & Nomoto 1986). While the synthesis of $^{60}$Co is negligible, the yield of $^{44}$Ti in the W7 Type Ia model is within a factor of two of the Type II/Ib models (see Fig. 1). However, core collapse supernovae occur about 6 times more frequently than thermonuclear supernovae in the Galaxy (van den Bergh & Tammann 1991; van den Bergh & McClure 1994). If all Type Ia events eject $1.8 \times 10^{-5}$ M$_\odot$ of $^{44}$Ti and the average Type II/Ib event ejects $3 \times 10^{-5}$ M$_\odot$, then standard deflagration Type Ia supernovae account for only $\sim 8\%$ of the Galactic $^{44}$Ca abundance. Some other deflagration models (Woosley & Eastman 1995) produce even less $^{44}$Ti, while still accounting for the majority of the spectral and light curve observations. It appears that standard Type Ia deflagration models do not make an appreciable fraction of the Galactic $^{44}$Ca abundance.

Several recent stellar evolution studies suggest an alternative progenitor for Type Ia supernovae, a 0.6 – 0.9 M$_\odot$ carbon–oxygen white dwarf that merges by gravitational radiation with a helium main-sequence star and accretes helium at a rate of several times $10^{-8}$ M$_\odot$ yr$^{-1}$ (Iben & Tutukov 1991; Limongi & Tornambe, 1991; Munari & Renzini, 1992; Tutukov, Yungelson & Iben 1992; Iben & Livio 1993; Tutukov & Yungelson 1995). When 0.15 – 0.20 M$_\odot$ of helium has been accreted in this model, a helium detonation is initiated at the base of the accreted layer. This helium detonation compresses the carbon-oxygen core and triggers a detonation there as well (Nomoto 1980, 1982; Woosley, Weaver & Taam 1980; Woosley, Taam & Weaver 1986; Livne 1990; Livne & Glasner 1991; Woosley & Weaver 1994; Woosley & Eastman 1995). The iron group nucleosynthesis is acceptable in these models (0.3 – 0.9 M$_\odot$ of $^{56}$Fe ejected), with $2.9 \times 10^{-4} - 3.9 \times 10^{-3}$ M$_\odot$ of $^{44}$Ti ejected. Production factors for $^{44}$Ca relative to its solar mass fraction thus range from 200 – 3000. The isotopes $^{46,47,48}$Ti and $^{51}$V also have large production factors in these models. The synthesis of $^{44}$Ti in these models is by a high temperature variety of incomplete explosive helium burning. Since the material never achieves NSE, it is not technically an alpha-rich freeze out. Depending upon how frequently these types of sub–Chandrasekhar mass white dwarfs explode, the large production factors suggest that these types of thermonuclear events might be the principal origin of $^{44}$Ca in nature. Sub-Chandrasekhar models also make significant amounts of $^{47}$Ti and $^{51}$V, which are also deficient in the stellar-chemical evolution models.

## 3. Detection of $^{44}$Ti

The radioactive isotope $^{44}$Ti decays to $^{44}$Sc with an accompanying emission of $\gamma$–rays at 67.85 and 78.38 keV. The most recent measurement of the $^{44}$Ti half–life gives a preliminary value of 58 years with a 1 sigma error of 10 years (Meißner et al. 1995). These values cover the range (46.4 to 66.6 yr) of previously reported values (Wing et al. 1965; Moreland & Heymann 1965; Frekers et al. 1983; Alburger & Harbottle 1990). The sources of the large uncertainty in the ongoing Meißner et al. (1995) measurement appear to be well understood and the uncertainty should drop in the near future to well below 10%. The isotope $^{44}$Sc then decays to the stable isotope $^{44}$Ca with a half–life of 5.7 hr, emitting a 1.157 MeV $\gamma$–ray. This line has been a recognized target for $\gamma$–ray astronomy for quite some time (e.g., Clayton et al. 1969)

Direct detection of $^{44}$Ti is possible through $\gamma$–ray (or hard x-ray) observations of individual supernova remnants or small groups of extinguishing OB associations. Additional means for inferring the abundance of $^{44}$Ti produced in supernovae are the analysis of late time supernova light curves and graphite grains from meteorites. In the remainder of this section we examine the evidence from each of these lines of inquiry in view of the production levels described in §2.

### 3.1 $^{44}$Ti in the Galaxy

Since the half–life of $^{44}$Ti is comparable to the average time between Galactic supernovae, its $\gamma$–ray emission should appear as a small collection of Galactic point sources. Of the six known local supernovae (which probably comprises a complete sample within a distance of $\simeq$ 3 kpc) only SN 1006 is considered to have been a Type Ia event, Cas A and Tycho were most likely Type Ib supernova, while the Crab, SN 1181 and Kepler were probably Type II supernovae. Of these six, only the $^{44}$Ti abundance in Cas A has been pursued aggressively and is discussed in §3.2. Evidence for Galactic $^{44}$Ti emission was sought by the large field of view spectrometers aboard the *High Energy Astronomy Observatory* (HEAO) and *Solar Maximum Mission* (SMM) satellites. Neither search for young, unknown and visually obscured supernovae remnants produced a positive $^{44}$Ti detection. Mahoney et al. (1992) estimated a 99% confidence level upper limit of $2 \times 10^{-4}$ photons cm$^{-2}$ s$^{-1}$ for the 67.85 and 78.38 keV line fluxes using the HEAO survey. Leising & Share (1994) placed a 99% confidence level upper limit of $8 \times 10^{-5}$ photons cm$^{-2}$ s$^{-1}$ for the 1.157 MeV line flux in the general direction of the Galactic center by analyzing 10 years of SMM data.

Detailed stellar-chemical evolution models of the Galaxy (e.g., Timmes et al. 1995a) produce only about 1/3 of the solar $^{44}$Ca abundance, with most of the production attributable to Type II supernova. This model provides satisfactorily reproduces the solar abundances for the vast majority of isotopes lighter than gallium, dwarf star elemental abundance histories, estimates of the present average Galactic supernova rates, spatial distribution and flux levels expected from $^{26}$Al and $^{60}$Fe (Hoffman et al. 1995; Timmes et al. 1995b) and the present interstellar medium $^{26}$Al/$^{27}$Al ratio. Note that the sum of $^{44}$Ca made as itself and $^{44}$Ti synthesized were used in the calculation of the solar $^{44}$Ca abundance. The mass of $^{44}$Ca produced as itself is comparable to the $^{44}$Ca synthesized as radioactive $^{44}$Ti.

To explain the solar $^{44}$Ca abundance, either the Type II $^{44}$Ti yields from Woosley & Weaver (1995) are too small (e.g., because of too much fall back in the more massive stars), the calculated supernova rates are too small, or there is another source of $^{44}$Ca in the Galaxy. Since $^{44}$Ti is produced in the innermost ejecta of the supernova, it is strongly dependent on the explosion details. This delicate dependence can be seen in Fig. 1 with the somewhat irregular behavior on progenitor mass. Explosions that one may regard as overly energetic, and neglecting any fallback material, alleviates most of the factor of 3 discrepancy (see Table 1). The iron yields in this case are, of course, much larger and may pose a problem for duplicating the abundance histories observed in dwarf stars.

Thielemann, Nomoto & Hashimoto (1996) have also examined the detailed nucleosynthesis of core collapse supernovae, and they find larger amounts of $^{44}$Ti is ejected. Their 13 $M_\odot$ star ejects 2.4 - 5.0 × $10^{-5}$ $M_\odot$ of $^{44}$Ti, a 15 $M_\odot$ star 7.2 - 12.0 × $10^{-5}$ $M_\odot$, a 20 $M_\odot$ star 17.0 × $10^{-5}$ $M_\odot$, and a 25 $M_\odot$ star 2.1 × $10^{-5}$ $M_\odot$. These values are sometimes similar, sometimes not, to those given in Table 1. Some of the differences between the Woosley & Weaver and Thielemann et al. yields are caused differences in the nuclear reaction rates. Hoffman et al. (1996) show that use of the Thielemann et al. reaction rates in the Woosley & Weaver stellar models produce nearly the same yields for the vast majority of isotopes. For the specific case of $^{44}$Ca, Hoffman et al. (1996) show a factor of 2 spread between the two groups due to differences in the relevant reaction rates.

Another reason for the differences in the yields is tied to the different adiabatic paths followed in the explosion, since the progenitor structure is quite similar between the two groups (Hoffman et al 1996). Thielemann et al. explode their stellar models by depositing thermal energy deep in the neutronized core (which later becomes the neutron star). The energy deposited gives a larger entropy to the innermost zones than what a piston driven explosion would impart and, in principle, a thermal explosion would eject more material. A larger entropy ensures a more vigorous alpha–rich freeze out and, thus, a larger $^{44}$Ti production. However, Thielemann et al. sum the ejecta from the outside in and place the mass cut (which is artificial in all models) at the position where sufficient $^{56}$Ni is produced to explain the observations. Woosley & Weaver, on the other hand, place the piston at a suitable mass cut and then follow the explosion. It is the difference between injecting momentum or energy in modeling the explosion that leads to the two groups following different adiabatic paths (also see Aufderheide, Baron & Thielemann 1991). What real self-consistent explosion models do, exploding via neutrino heating and multidimensional convection, might be reflective of one dimensional momentum driven explosions, one dimensional energy driven explosions, or some intermediate case. Hence, the differences between the two groups in the amount of $^{44}$Ti ejected is an example of the spread one obtains due to uncertainties in modeling the Type II explosion mechanism.

A leading candidate for an additional source of $^{44}$Ti would be the sub-Chandrasekhar mass white dwarf models for Type Ia supernovae discussed in the previous section or other Type Ia models where a detonation wave passes through a surface layer of helium. Nature may also prefer sub-Chandrasekhar mass Type Ia supernovae to make the isotopes $^{47}$Ti and $^{51}$V, which are also deficient by a factor of 3 in the stellar-chemical evolution models. A larger $^{44}$Ti abundance from any of these three possibilities (larger yields, larger supernova rate, additional source) could account for a significant fraction of the strong diffuse Galactic 511 keV emission due to electron–positron annihilation (Bussard, Ramaty, & Drachman 1979; Woosley & Pinto 1988; Lingenfelter, Chan, & Ramaty 1993).

### 3.1.1 The diffuse Galactic annihilation radiation

Observations of the 511 keV line from positron annihilation show a steady diffuse component from the Galactic disk superimposed upon a time–variable point source located near the Galactic center (Johnson, Harden & Haymes 1972; Leventhal, MacCallum & Stang 1978; Niel et al. 1990; Gehrels et al. 1991; Skibo, Ramaty, & Leventhal 1992; Leventhal et al. 1993; Purcell et al. 1993; Ramaty, Skibo. & Lingenfelter 1994). Decay products from radioactive decay of $^{26}$Al, $^{44}$Ti and $^{56}$Co produced by supernovae, novae or the winds of Wolf-Rayet stars have been proposed as a source for the diffuse component (Ramaty & Lingenfelter 1979; Woosley & Pinto 1988; Lingenfelter et al. 1993), along with smaller contributions from the decay of pions produced by cosmic ray interactions (Ramaty, Kozlovsky & Lingenfelter 1979; Ramaty & Lingenfelter 1991), energetic solar flares (Mandzhavidze & Ramaty 1992), pulsars (Sturrock 1971) and gamma-ray bursts (Lingenfelter & Hueter 1984). However, none of the previous estimates for the contribution from radioactive nuclei produced in supernova enjoy the number and/or quality of the stellar models employed here.

Two positrons are produced in 95% of the decays from $^{44}$Ti to $^{44}$Ca, while 85% of the decays from $^{26}$Al to $^{26}$Mg produce a single positron. That portion of pair annihilation radiation due to the decay of $^{44}$Ti and $^{26}$Al balance the steady state production of these isotopes. Once the positrons are in the interstellar medium, they have a mean life against annihilation of at least $10^5$ yr (Bussard et al. 1979; Lingenfelter et al. 1993; Chan & Lingenfelter 1994; Ramaty et al. 1994). This is much larger than the average time interval between supernovae, so the steady state assumption is valid.

For the exponential disk + $1/r^2$ bulge Galactic mass distribution used by Timmes et al. (1995a) and requiring a full production of the solar abundance of $^{44}$Ca, we calculate annihilation rates of $2.9 \times 10^{42}$ positrons s$^{-1}$ due to $^{26}$Al, $4.5 \times 10^{42}$ positrons s$^{-1}$ due to $^{44}$Ti, and $1.6 \times 10^{43}$ positrons s$^{-1}$ from $^{56}$Co. The total annihilation rate is then $2.3 \times 10^{43}$ positrons s$^{-1}$ with 13 % of the 511 keV emission explained by the decay of $^{26}$Al in the interstellar medium. Lingenfelter et al. (1996), using a similar mass distribution, estimated a total annihilation rate of $1.5 \pm 0.5 \times 10^{43}$ positrons s$^{-1}$, while measurements of the diffuse 1809 keV $^{26}$Al $\gamma$-ray line emission constrain the positron contribution due to $^{26}$Al to $16 \pm 5\%$ (Mahoney et al. 1982, 1984; Share et al. 1985; Gehrels et al. 1993; Skibo et al 1992; Diehl et al. 1995)

In this simple picture, the $^{44}$Ti yields from Type Ia and Type II/Ib were artificially increased by a factor of 3 so that a solar $^{44}$Ca abundance was obtained at a time 4.6 Gyr ago at a distance 8.5 kpc from the Galactic center (see the discussion in §3.1). The supernova rates calculated by the stellar-chemical evolutions were 3.1 per century for Type II + Ib supernova and 0.53 per century for Type Ia supernova, in good agreement with the estimates of van den Bergh & Tammann (1991) and van den Bergh & McClure (1994). The annihilation rate due to $^{56}$Co assumed an average escape fraction of 3%. However, $^{56}$Co has a short mean lifetime of 111.4 days and positrons are only produced in 19% of the $^{56}$Co decays so that a large fraction could annihilate inside the supernova ejecta before the mean free path becomes long enough for the positrons to escape. In addition, a magnetic field of $10^{-6}$ gauss gives a gyroradius smaller than the homologous expansion radius at 111.4 days. Thus, it is possible that the escape fraction of positrons from $^{56}$Co is an order of magnitude smaller than the 3% used above.

In the limiting case that the escape fraction of positrons from the decay of $^{56}$Co is zero, and the $^{44}$Ti mass ejected by supernovae is pushed to the point of over-producing the solar $^{44}$Ca abundance by a factor of 2, the annihilation rate due to $^{44}$Ti increases to $10^{43}$ positrons s$^{-1}$. The total annihilation rate in this case is $1.3 \times 10^{43}$ positrons s$^{-1}$ with 23% emission explained by the decay of $^{26}$Al. The error bars on the total annihilation rate and the $^{26}$Al contribution (unresolved angular distribution, subtraction of the time-variable contribution, amount of discrete foreground source contamination) as well as the uncertainties in the supernova models thus allow the possibility that a significant fraction, if not most, of the smooth 511 keV emission is from the annihilation of $^{44}$Ti positrons.

## 3.2 $^{44}$Ti in Cas A

The Cas A supernova remnant is relatively close ($\simeq$ 3 kpc), young ($\simeq$ 300 yr) and wide (physical diameter $\simeq$ 4 pc), making it one of the premier sites for studying the composition and early behavior of a supernova remnant. It may only be equaled as SN 1987A unfolds. During July 1992 and February 1993 Cas A was observed ($l$=111.73°, $b$=-2.13°) with the COMPTEL telescope aboard the *Compton Gamma-Ray Observatory* (CGRO), and showed evidence for line emission at 1.157 MeV from the decay of $^{44}$Ti at a significance level of $\simeq 4\sigma$ (Iyudin et al. 1994). A total of 330 photons were attributed to the 1.157 MeV feature. Adopting a distance of 2.8 kpc, the initially measured line flux of $7.0 \pm 1.7 \times 10^{-5}$ photons cm$^{-2}$ s$^{-1}$ (detection limit of $10^{-5}$ photons cm$^{-2}$ s$^{-1}$) converts into a $^{44}$Ti mass ejected during the Cas A supernova explosion of $1.4 - 3.2 \times 10^{-4}$ M$_\odot$, depending on the precise date of the event and the mean lifetime of $^{44}$Ti (Iyudin et al. 1994). This was the first (and so far only) time a positive detection of $\gamma$-ray emission from the decay of $^{44}$Ti has been reported, either in the debris left by individual supernova or through point sources of Galactic emission.

A recent analysis of additional COMPTEL observations of Cas A were unable to attribute a comparable number of photons to the 1.157 MeV feature, giving a lower flux of $4.2 \pm 0.8 \times 10^{-5}$ photons cm$^{-2}$ s$^{-1}$ (Schönfelder et al. 1995), corresponding to slightly smaller $^{44}$Ti mass of $1 - 2 \times 10^{-4}$ M$_\odot$.

During July 1992 and February 1993 the Galactic plane in the region near Cas A was also observed with OSSE CGRO, and showed no firm evidence of emission from the three strongest decay lines (67.85 keV, 78.38 keV and 1.157 MeV) of $^{44}$Ti (The et al. 1995). That is, OSSE did not confirm the COMPTEL detection of the $^{44}$Ti decay lines. From simultaneous fits to the three $\gamma$-ray line fluxes and reasonable estimates of the systematic errors, The et al. (1995) derive a 99% confidence upper limit of $5.5 \times 10^{-5}$ photons cm$^{-2}$ s$^{-1}$ to the flux in each line. Adopting a distance of 2.92 kpc to the Cas A remnant, the upper limit line flux corresponds to $1.7 - 5.0 \times 10^{-4}$ M$_\odot$ of $^{44}$Ti ejected during explosion, depending again on the age of Cas A and the $^{44}$Ti mean lifetime. Considering the statistical uncertainties only, the COMPTEL (Iyudin et al. 1994) and OSSE (The et al. 1995) $^{44}$Ti flux measurements are formally consistent with each other only at the 1.8% confidence level.

More recently however, OSSE observations of Cas A (The et al. 1996) have tripled their exposure time. The best simultaneous fit to the flux of the three $^{44}$Ti lines yields $1.7 \pm 1.4 \times 10^{-5}$ photons cm$^{-2}$ s$^{-1}$ (The et al. 1996). Thus, the total COMPTEL and OSSE observations are now less in conflict, and may be converging on a lower value than COMPTEL originally reported. At the $1\sigma$ level the two experiments are in agreement near $3.5 \times 10^{-5}$ cm$^{-2}$ s$^{-1}$, implying $1.0 \times 10^{-4}$ M$_\odot$ of $^{44}$Ti.

The larger values of $^{44}$Ti ejected during explosion that are inferred from the COMPTEL observations would be difficult to accommodate in the supernova models; the smaller ones might not (see Figure 1). Hoffman et al. (1995) have pointed out that a large mass of $^{44}$Ti cannot be ejected unless a large quantity of $^{56}$Ni is also ejected, and this implies a bright supernova. How bright? An abundance of $^{44}$Ti as large as reported ($\sim 10^{-4}$ M$_\odot$) would probably mean the ejection of at least 0.05 M$_\odot$ of $^{56}$Ni and possibly considerably more (Woosley et al. 1995). With or without a hydrogen envelope, the supernova would then have a peak luminosity brighter than the $10^{42}$ erg s$^{-1}$ observed for SN 1987A. For a distance of 2.8 kpc to Cas A, this would imply a peak absolute magnitude of -4, assuming no interstellar absorption. Cas A was certainly not detected as such a bright object. Ashworth (1980) has argued for a possible eyewitness observation of the Cassiopeia supernova by John Flamsteed (first astronomer royal of England) in 1680 when the new star had an apparent brightness of 6th magnitude. The absence of widespread reports of the event probably means that the peak visual magnitude was fainter than 3rd magnitude (van den Bergh 1991; Da Silva 1993). A peak absolute magnitude of -4 certainly would have been recorded by both Eastern and Western observers.

However, there may be large reddening corrections. Minkowski (1957) and Searle (1971) estimate a visual extinction of 4.3 magnitudes in the direction of Cas A, while Peimbert & van den Bergh (1971) suggest a visual reddening correction of 5 – 7 magnitudes due to the circumstellar material surrounding the supernova remnant. If Cas A was embedded in a dusty region or experienced significant mass loss which condensed into dust grains before the explosion, the extinction could have been even larger. Recent measurements of the X-ray scattering halo around Cas A (Predehl & Schmitt 1995) from the Röntgen satellite (ROSAT) offer some compelling evidence for a larger reddening correction. The ROSAT measurements show that the interstellar absorption column depth of N$_H$ is a factor of $\sim 2$ larger than the column depth of N$_H$ that normally corresponds to an optical extinction of A$_V$ = 5. The material causing the N$_H$ excess appears to be metal–rich, dust poor and distributed close (relative to its distance) to the remnant (Predehl & Schmitt 1995). If this material had a typical gas/dust ratio at the time of the Cas A explosion, then an additional visual extinction of 5 magnitudes may apply. A total visual extinction of 10 magnitudes is near the value needed to reconcile the Flamsteed detection and our estimate of the peak magnitude based on nucleosynthesis arguments. It is thus conceivable that a dusty shell prevented the widespread detection of this supernova, and that the supernova shock wave destroyed the dust as it propagated through the debris and the material surrounding the Cas A supernova (Hartmann et al. 1996).

A progenitor more massive than 10 $M_\odot$ has been suggested for Cas A by analysis of the fast moving optical knots in the supernova remnant (Brecher & Wasserman 1980; Ellison et al. 1994), while estimates of the mass emitting X–rays range from 2.4 – 15.0 $M_\odot$ (Fabian et al. 1980; Tsunemi et al. 1986; Jansen et al. 1988; Ellison et al. 1994). The three-dimensional structure of the remnant has been explored (Reed et al. 1995). These estimates suggests that if Cas A was a Type II event, the progenitor mass was between 12 and 25 $M_\odot$. Combined with the COMPTEL/OSSE observations of $\sim 10^{-4}$ $M_\odot$ of $^{44}$Ti being ejected, implies that there was little or no fall back during the explosion. The actual mass at the time of the explosion is critical. A 20 - 25 $M_\odot$ supernova might eject $\sim 10^{-4}$ $M_\odot$ of $^{44}$Ti and still leave a remnant of baryonic mass 2.0 $M_\odot$ that might become a black hole (Woosley & Weaver 1995). A presupernova lighter than 19 $M_\odot$ would almost certainly leave a neutron star (Timmes, Woosley, & Weaver 1996). A Type Ib progenitor for Cas A (e.g., Fesen, Becker & Blair 1987; van den Bergh 1991; Fesen & Becker 1991) even if it had a large main sequence mass, might have lost sufficient mass as a Wolf-Rayet star to leave a neutron star remnant. Thus the nature of the compact remnant in Cas A, from a theoretical view, is ambiguous. Observationally, X-ray imaging does not reveal the synchrotron nebula one might expect around a neutron star (Murray et al. 1979; Pravado & Smith 1979; Becker et al. 1980; Jansen et al. 1988; Ellison et al. 1994; Hartmann et al. 1996). Deep images taken under excellent seeing conditions have set strict magnitude limits ($M_I \gtrsim 8.9$, $M_R \gtrsim 9.3$) on the presence of a stellar remnant (Van Den Bergh & Pritchet 1986). The region around the center of Cas A simply appears void of any detectable stars. So observationally, a black hole may have formed. This would make the ejection of a lot of $^{44}$Ti more difficult, but not impossible. The black hole mass would need to be not too far above the critical mass, probably less than 2.0 $M_\odot$ (gravitational mass).

### 3.3 $^{44}$Ti in SN 1987A

One of the advantages of a nearby supernova like SN 1987A is the opportunity to follow the light curve and spectrum to much later times than has been possible in any other event. An important effect of $^{44}$Ti in SN 1987A is to provide a floor to the bolometric light curve that is proportional to the uncertain abundance produced in the explosion (Pinto & Woosley 1988; Woosley, Pinto, & Hartmann 1989; Kumagai et al. 1989; Arnett et al. 1989). The thermal luminosity deriving from $^{44}$Ti decay to $^{44}$Ca is

$$L_{44} = 4.1 \times 10^{36} \left[1 - \exp\left(-\kappa_{44}\phi_0(t_0/t)^2\right) + 1.3\right] \times$$
$$\exp\left(-t/\tau_{44}\right) \left(\frac{\mathrm{M}(^{44}\mathrm{Ti})}{1.0 \times 10^{-4}}\right) \mathrm{erg\ s}^{-1} , \qquad (1)$$

where $\phi_0 = 7.0 \times 10^4$ g cm$^{-2}$ is the column depth at some fiducial time $t_0 = 10^6$ s, and $\kappa_{44} = 0.04$ cm$^2$ g$^{-1}$ is an estimate of the effective opacity (Pinto & Woosley 1988; Woosley, Pinto & Hartmann 1989).

Observed and theoretical bolometric light curves of SN 1987A are shown in Fig. 3a for the 500 to 1500 day time period and in Fig. 3b for the 500 to 3500 day period. In each case the observed bolometric luminosity (e.g., Suntzeff et al. 1992) is shown as the filled circles. There are no data points past day 2000 since most the emission is in the far infrared wavelength region where it is not easily observed. A rather large, but not unphysical, ejected mass of M($^{44}$Ti) = $1.0 \times 10^{-4}$ M$_\odot$ was used. Changes in the amount of $^{44}$Ti ejected produce a linear shift of the light curve (see equation 1). The labeled dashed lines indicate the $\gamma$–ray luminosity of the various radioactive isotopes and the solid line is the total luminosity, assuming that radioactive decay is the sole power source. Radioactive $^{44}$Ti tends to dominate prompt radioactive contributions to the bolometric light curve after about 1500 days due to the half–life of $^{44}$Ti and local deposition of the positron kinetic energy ($^{44}$Sc to $^{44}$Ca). Figure 4 shows the small effect of the uncertainty in the effective opacity $\kappa_{44}$. Factors of two variation from the value used above are shown as dashed lines. The figures suggest that we are now entering an epoch in the supernova's life when the dominant energy source, exclusive of input from a pulsar, accreting compact object or circumstellar interaction, should be the decay of radioactive $^{44}$Ti with a small current contribution from $^{60}$Co (discussed in §4). This energy input may be difficult to uniquely resolve, however, if the atomic processes that convert nuclear decay energy into optical-infrared luminosity are no longer operating in steady state (Clayton et al. 1992; Fransson & Kozma 1993).

For a distance of 50 kpc to SN 1987A and a half–life of 58 yr, $5 \times 10^{-5}$ M$_\odot$ of $^{44}$Ti would produce a $\gamma$–ray line flux of $\simeq 1 - 2 \times 10^{-6}$ photons cm$^{-2}$ s$^{-1}$. This line flux is too small for CGRO and probably too small for *International Gamma-Ray Astrophysics Laboratory* (INTEGRAL) instruments, but large enough that it might be detected in the next century by post-INTEGRAL experiments.

### 3.4    $^{44}$Ti in Meteorites

Interstellar silicon carbide isolated from the Murchison carbonaceous meteorite is anomalous in its C, Si, Mg, Ti, Sr, Ba, Nd and noble gas isotopic compositions (Zinner, Tang & Anders 1989; Ott & Begemann 1990; Lewis, Amari & Anders 1990, 1994; Hoppe et al. 1994; see also Anders & Zinner 1993). Most silicon carbide grains carry the signature of hydrogen and helium burning with s-process trace elements. A few of the grains, however, exhibit extremely exotic isotopic compositions, distinct from the majority of the silicon carbide grains while being indistinguishable in their morphology. These grains were named "type X" grains (Zinner et al. 1991; Amari et al. 1992) and show large excesses of $^{12}$C relative to $^{13}$C (up to 76 times solar), $^{15}$N relative to $^{14}$N (up to 22 times solar), $^{26}$Al/$^{27}$Al ratios up to 0.6, depletion in $^{29,30}$Si relative to $^{28}$Si (up to 77%), and $^{49}$Ti excesses up to 95% relative to $^{48}$Ti (Amari et al. 1992; Nittler et al. 1996; Hoppe et al. 1996; Nittler 1995 private communication). Ion micro-probe mass spectrometer analyses of one X type grain from the Murchison carbonaceous meteorite showed a large $^{44}$Ca excess (relative to $^{40}$Ca) of 300% (Amari et al. 1992). New (unpublished) data on another 9 silicon carbide X type grains and a few graphite grains have inferred $^{44}$Ti/$^{48}$Ti ratios which range from 0.0033 to 0.058 (Amari et al. 1995; Nittler et al. 1995), while in another silicon carbide grain this ratio is an astonishingly large 0.37 (Hoppe et al. 1996).

Assuming that grains form in supernovae, the Ca and Ti isotopic anomalies confirm that production of $^{44}$Ca occurs during explosive nucleosynthesis in supernovae. (Or as the meteoritic community prefers, assuming supernovae produce $^{44}$Ca the isotopic anomalies strongly suggests grain condensation from supernova ejecta.) The requirement that $^{44}$Ti, produced in the deepest layers of the supernova to be ejected, be present in an environment that is carbon rich also requires considerable mixing of different zones prior to or during the grain formation epoch since carbon exceeds oxygen only in the helium shell. Large $^{44}$Ca anomalies had been predicted by Clayton (1975) based on the condensation of radioactive $^{44}$Ti. Clayton envisioned production of $^{44}$Ti in explosive helium, oxygen, and silicon burning. It is synthesized therein, sufficiently for production of $^{44}$Ca excesses in objects having an enhanced Ti/Ca ratio, but those zones do not provide the bulk of $^{44}$Ti synthesized in massive stars. It came as a major surprise to find these titanium anomalies in carbide grains, since none of the explosive zones in question contain significant amounts of carbon. While it has recently been realized that the ejecta from supernova are subject to a variety of macroscopic dynamical instabilities (chiefly Rayleigh-Taylor and Richtmyer-Meshkov) that induce mixing between zones (e.g., Herant & Woosley 1994), the need for mixing at a microscopic level is still surprising. Until calculations are done that include a realistic description of the turbulent cascade to small scales and until our understanding of grain chemistry in young supernovae advances further, we are unable to draw quantitative conclusions regarding the amount of $^{44}$Ti required to explain these observations.

## 4. Detection of $^{60}$Co

Each beta decay of $^{60}$Co to $^{60}$Ni (half–life of 5.27 years; Tuli 1990) produces $\gamma$–rays of 1.332 MeV and 1.173 MeV for a total $\gamma$–ray energy of 2.505 MeV. The maximum kinetic energy of the emitted electron is $Q = 0.3178$ MeV. We assume that the electron deposits its energy locally even when the gamma-rays escape. Lacking a detailed Kurie plot for this nucleus, an estimate of the average kinetic energy of the neutrino is given by the approximate formula

$$\langle E_\nu \rangle \simeq \left(\frac{Q'}{2}\right)\left(1 - \frac{1}{W^2}\right)\left(1 - \frac{1}{4W} - \frac{1}{9W^2}\right) \tag{2}$$

where

$$W = 1 + \frac{Q}{m_e c^2} = \frac{Q'}{m_e c^2} . \tag{3}$$

The average electron kinetic energy is then the maximum kinetic energy $Q$ minus the average neutrino kinetic energy. One concludes that the beta decay electron contributes about 7% of the $\gamma$–ray energy from $^{60}$Co. Since the electron's kinetic energy is deposited locally, however, it remains an energy source in the supernova remnant for several half–lives of $^{60}$Co.

At present, detection of $^{60}$Co is limited to Galactic point source $\gamma$–ray surveys and the analysis of late time supernova light curves. Leising & Share (1994) searched almost 10 yrs of data from the SMM Gamma-Ray Spectrometer for evidence of $\gamma$-ray line emission from the decay of $^{60}$Co. Although Leising & Share found no direct evidence of emission, they placed a 99% confidence level upper limit of 8 × 10$^{-5}$ photons cm$^{-1}$ s$^{-1}$ for the 1.173 MeV line integrated over the central radian of Galactic longitude. Subsequent analysis of this result suggested an upper limit to the mass of 1.7 M$_\odot$ of $^{60}$Fe in the interstellar medium today, which is close to what stellar-chemical evolution calculations yield (Timmes et al. 1995b).

### 4.1 $^{60}$Co in SN 1987A

In old supernova remnants the decay of $^{60}$Co occurs at the same rate as the $^{60}$Fe decay that feeds it while in young remnants, like SN 1987A, $^{60}$Co needs to be considered independently. The $\gamma$–ray luminosity from the beta decay of $^{60}$Co is

$$L_{60} = 6.7 \times 10^{36} \left[1 - \exp\left(-\kappa_{60}\phi_0(t_0/t)^2\right) + 0.42\right] \times$$

$$\exp\left(-t/\tau_{60}\right) \left(\frac{\text{M}(^{60}\text{Co})}{2.0 \times 10^{-5}}\right) \text{ erg s}^{-1} , \tag{4}$$

where $\phi_0 = 7.0 \times 10^4$ g cm$^{-2}$ is the column depth at some fiducial time $t_0 = 10^6$ s, and $\kappa_{60} = 0.04$ cm$^2$ g$^{-1}$ is an estimate of the effective opacity (e.g., Woosley, Pinto & Hartmann 1989).

The $\gamma$–ray deposition efficiency (first term in equation 4) is a key uncertainty. Both the column depth and effective opacity are uncertain due to mixing and clumping. A more rigorous light curve calculation would distribute the $^{60}$Co, perform Monte Carlo calculations and average over scattering angles (Woosley, Pinto & Hartmann 1989, Woosley & Eastman 1994). In the simple approach used here, we absorb all these uncertainties into the value for $\kappa_{60}$ and examine the sensitivity of the main results given below by varying the effective opacity by a factor of two (in addition to the uncertainty associated with the yield of $^{60}$Co itself).

Figs. 3a and 3b show the effects expected on the bolometric light curve of SN 1987A for an ejected mass of M($^{60}$Co) = $2.0 \times 10^{-5}$ M$_\odot$. The figures suggest that $^{60}$Co could have been contributing appreciably to the light curve at 1500 days and might even contribute to the light curve at the 10% level at 3500 days, but the level of the effect is less than the substantial error bars on the current luminosity measurements. There may also be significant modifications to the late time light curve due to the freezing out of atomic processes and the failure of the steady state assumption (Clayton et al. 1992; Fransson & Kozma 1993). The effect of varying the canonical value of the effective opacity (dotted curve), $\kappa_{44}=\kappa_{60}=0.40$ cm$^2$ g$^{-1}$, by factors of two (dashed curves) is shown in the upper panel of Figure 4. The changes are larger for $^{60}$Co than $^{44}$Ti since $^{60}$Co deposits a larger energy per decay than $^{44}$Ti.

For a distance of 50 kpc to SN 1987A and a half–life of 5.27 yr, an ejected mass of M($^{60}$Co) = $2 \times 10^{-5}$ would produce a $\gamma$–ray line flux of $\simeq 3 - 5 \times 10^{-6}$ photons cm$^{-2}$ s$^{-1}$ at 1500 days. This line flux might be detectable by future gamma-ray astronomy missions, such as INTEGRAL, though the short half-life makes a detection of SN 1987A unlikely.

## 5. Summary

Recent grids of Type II and Type Ib models predict $^{44}$Ti yields of between $10^{-5}$ to $10^{-4}$ M$_\odot$, with typical values $\simeq 3 \times 10^{-5}$ M$_\odot$ for the Type II models and $\simeq 6 \times 10^{-5}$ M$_\odot$ for the Type Ib models. Producing more than $10^{-4}$ M$_\odot$ of $^{44}$Ti is difficult but making much less is easy since $^{44}$Ti is synthesized in a region very close to the mass–cut. Yields of $^{60}$Co from core collapse supernovae vary between $10^{-6}$ and $10^{-4}$ M$_\odot$, with typical values $\simeq 3 \times 10^{-5}$ M$_\odot$ for the Type II models and $\simeq 2 \times 10^{-6}$ M$_\odot$ for the Type Ib models. Detailed chemical evolution models that use only Type II and Ib supernovae and Chandrasekhar mass models for Type Ia underproduce the solar abundance of $^{44}$Ca by a factor of 3. The production site of this isotope may be the explosion of sub-Chandrasekhar mass white dwarfs which have accreted a helium shell. There is the possibility that a significant fraction, if not most, of the steady state diffuse component of the 511 keV emission is from the annihilation of positrons due to the decay of $^{44}$Ti.

The line flux derived from COMPTEL and OSSE observations of Cas A give an upper bound for the mass of $^{44}$Ti ejected which is consistent with the core collapse models. The abundance of $^{44}$Ti suggested by the COMPTEL observations raises the possibility that Cas

A may have been a brighter supernova than previously thought. If Cas A was a Type Ib supernovae which experienced significant mass loss that condensed into dust grains prior to the explosion, the visual extinction could have been very large. Our estimates suggest that an extinction of 10 magnitudes is necessary to understand the historical limits on brightness. The abundances of $^{44}$Ti and $^{60}$Co predicted by Type II supernova models suggest the mass of $^{44}$Ti ejected by SN 1987A may be detectable by the next generation of $\gamma$-ray telescopes, and that $^{60}$Co can contribute to the late time light curves of SN 1987A and other Type II supernovae at the 10% level.

This work has been supported at Chicago by an Enrico Fermi Postdoctoral Fellowship (F.X.T); at Santa Cruz by the NSF (AST 91 15367), NASA (NAGW 2525), and the California Space Institute (CS86-92); and at Clemson by NASA (NAG 5-1578) and a Compton Gamma Ray Observatory Postdoctoral Fellowship (F.X.T).

# TABLE 1

Massive Star $^{44}$Ti Yields[a]

| Mass | External to Piston | Ejected |
| --- | --- | --- |
| 11 | 6.2E-5 | 6.2E-5 |
| 12 | 8.3E-5 | 8.3E-5 |
| 13 | 7.1E-5 | 6.3E-5 |
| 15 | 9.1E-5 | 5.7E-5 |
| 18 | 8.3E-5 | 8.5E-6 |
| 19 | 1.2E-4 | 1.9E-5 |
| 20 | 1.2E-4 | 1.4E-5 |
| 22 | 1.5E-4 | 6.1E-5 |
| 25 | 1.6E-4 | 3.0E-5 |
| 30 | 1.7E-4 | 1.4E-4 |
| 35 | 1.9E-4 | 1.8E-9 |
| 40 | 2.0E-4 | 1.5E-10 |

[a] All entries in solar masses for solar metallicity stars.

Figures and Captions

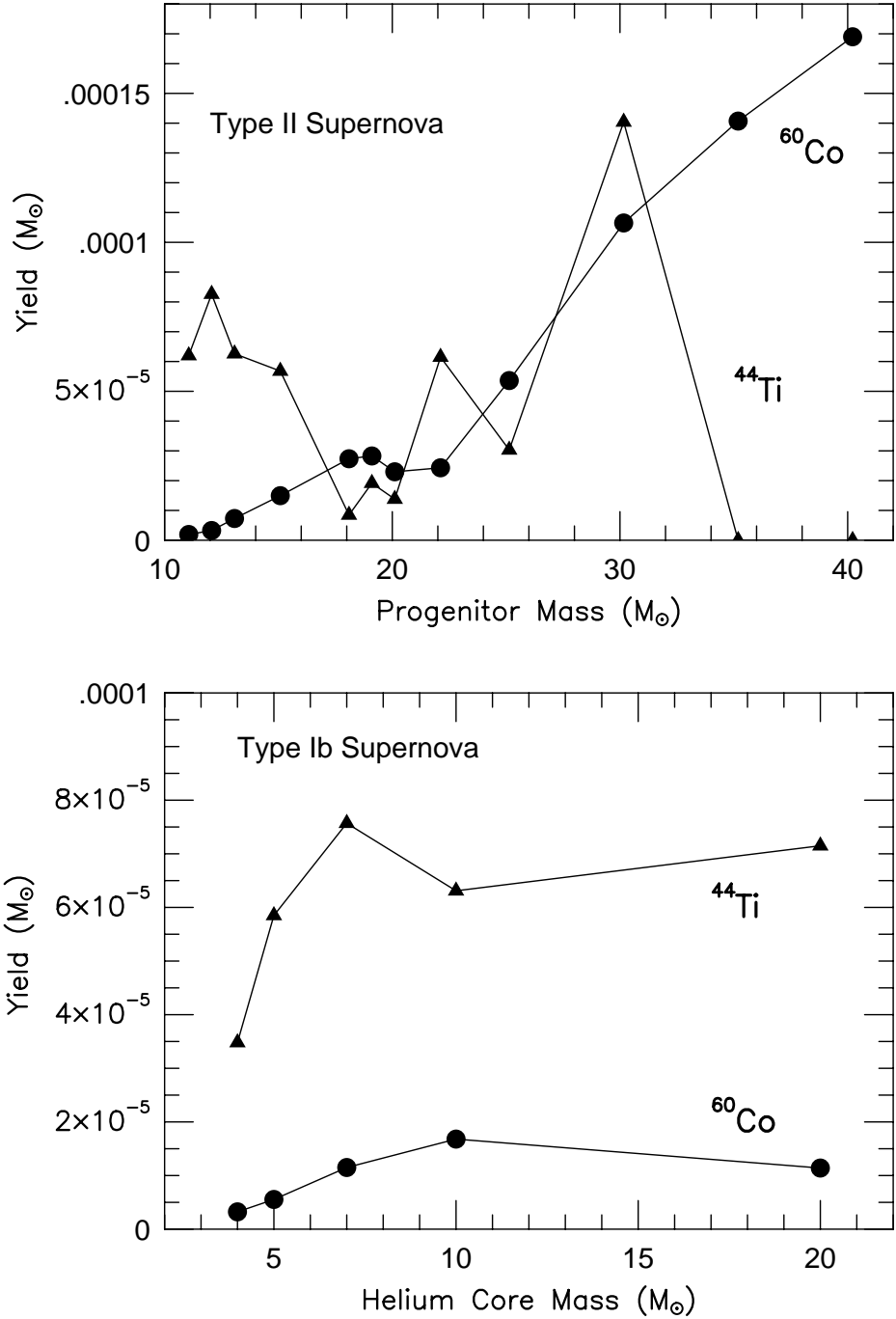

Fig. 1.— Yields of $^{44}$Ti and $^{60}$Co from the solar metallicity (a) Type II supernovae models of Woosley & Weaver (1995) and (b) Type Ib supernovae models of Woosley, Langer & Weaver (1995). The large non-monotonic variation in the $^{44}$Ti yields for the Type II supernovae shows the sensitivity to "fall back" for this species made in the innermost shell to be ejected. Synthesis of $^{60}$Co is less sensitive to this uncertainty.

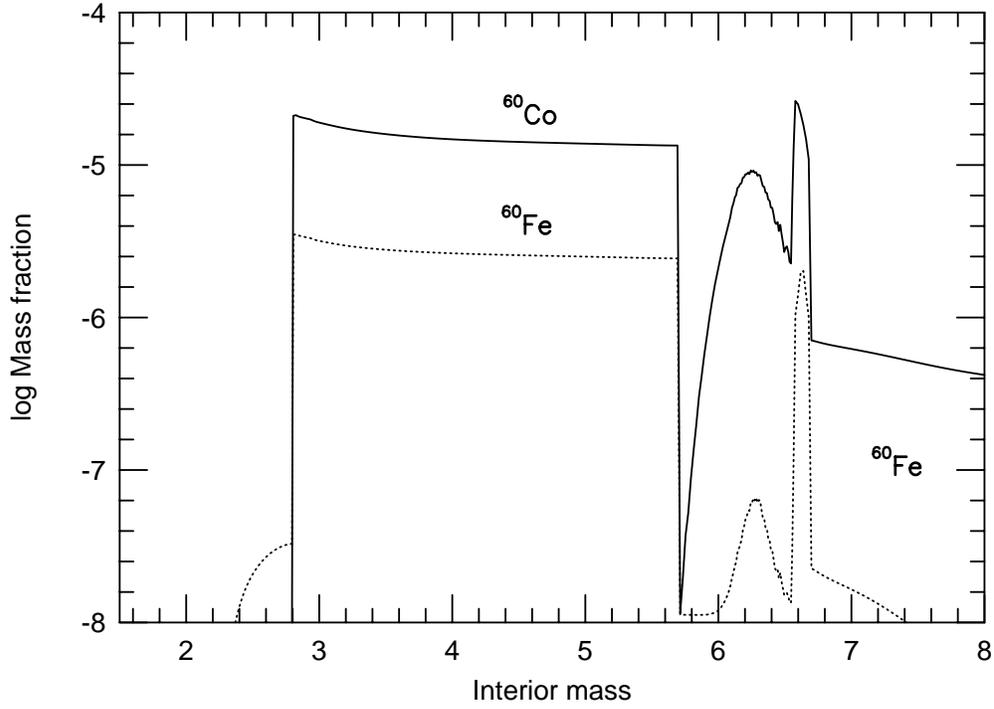
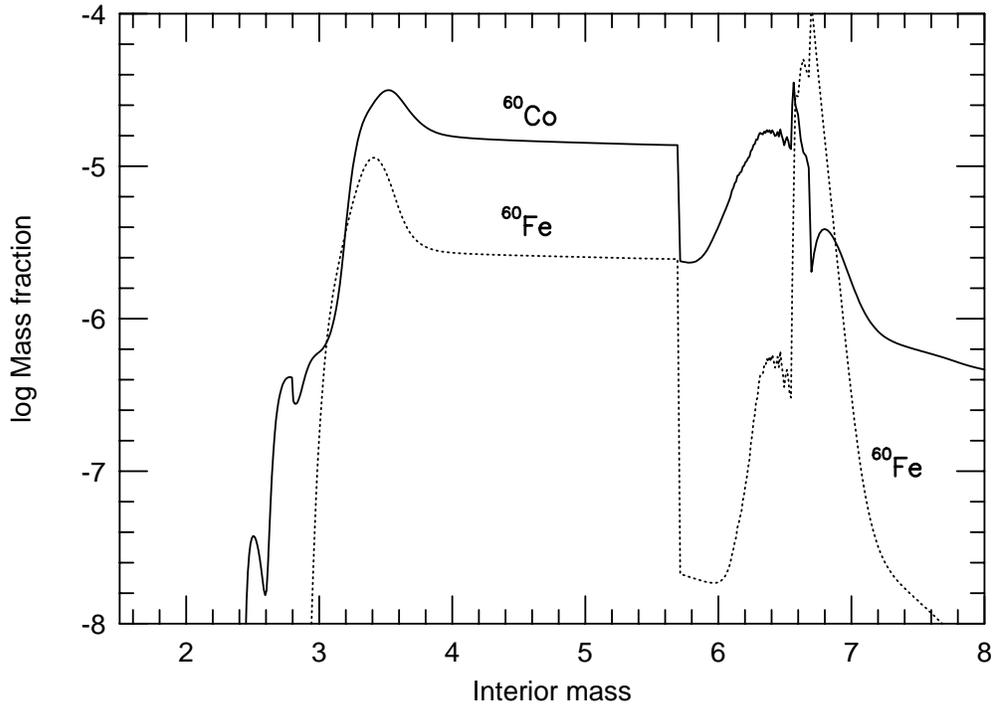

Fig. 2.— Abundance profiles of $^{60}$Co and $^{60}$Fe vs interior mass (M$_\odot$) for a solar metallicity, 25 M$_\odot$ (a) presupernova star and (b) exploded star at $2.5 \times 10^4$ s.

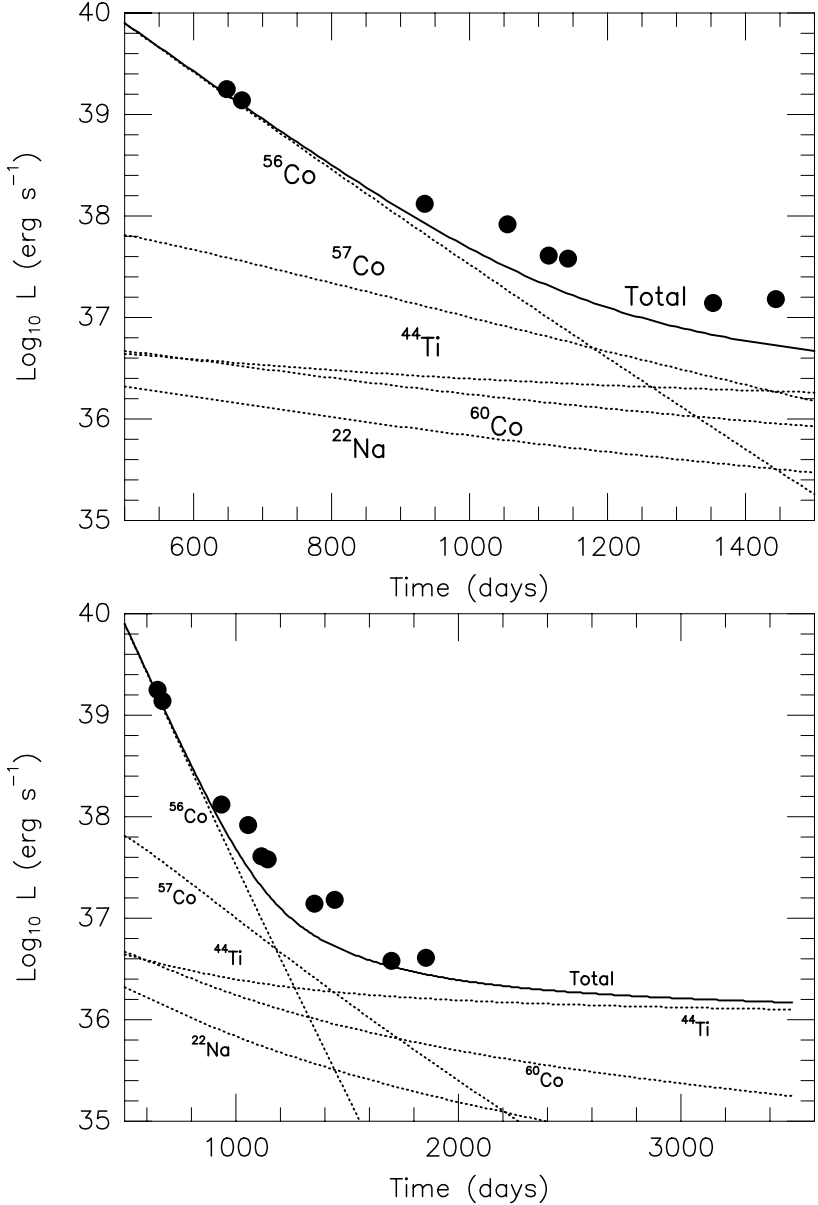

Fig. 3.— Bolometric light curve of SN 1987A from (a) 500 to 1500 days and (b) 500 to 3500 days. The observed bolometric luminosity is shown as the filled circles (Suntzeff et al. 1992). Contributions from the decay of $^{56,57}$Co, $^{22}$Na, $^{44}$Ti, and $^{60}$Co are shown by the labeled dashed lines while the solid line is the total calculated $\gamma$–ray luminosity. Ejected masses of M($^{44}$Ti) = 1.0 × 10$^{-4}$ M$_\odot$ and M($^{60}$Co) = 2.0 × 10$^{-5}$ M$_\odot$, along with effective opacities of $\kappa_{44} = \kappa_{60} = 0.04$ cm$^2$ g$^{-1}$ were used. The relatively long half–life and local deposition of the $^{44}$Sc to $^{44}$Ca positron kinetic energy, $^{44}$Ti tends to dominate the bolometric light curve after about 1500 days. However, $^{60}$Co might have been contributing appreciably to the light curve at 1500 days and may contribute to the bolometric light curve at about the 10% level at 3500 days.

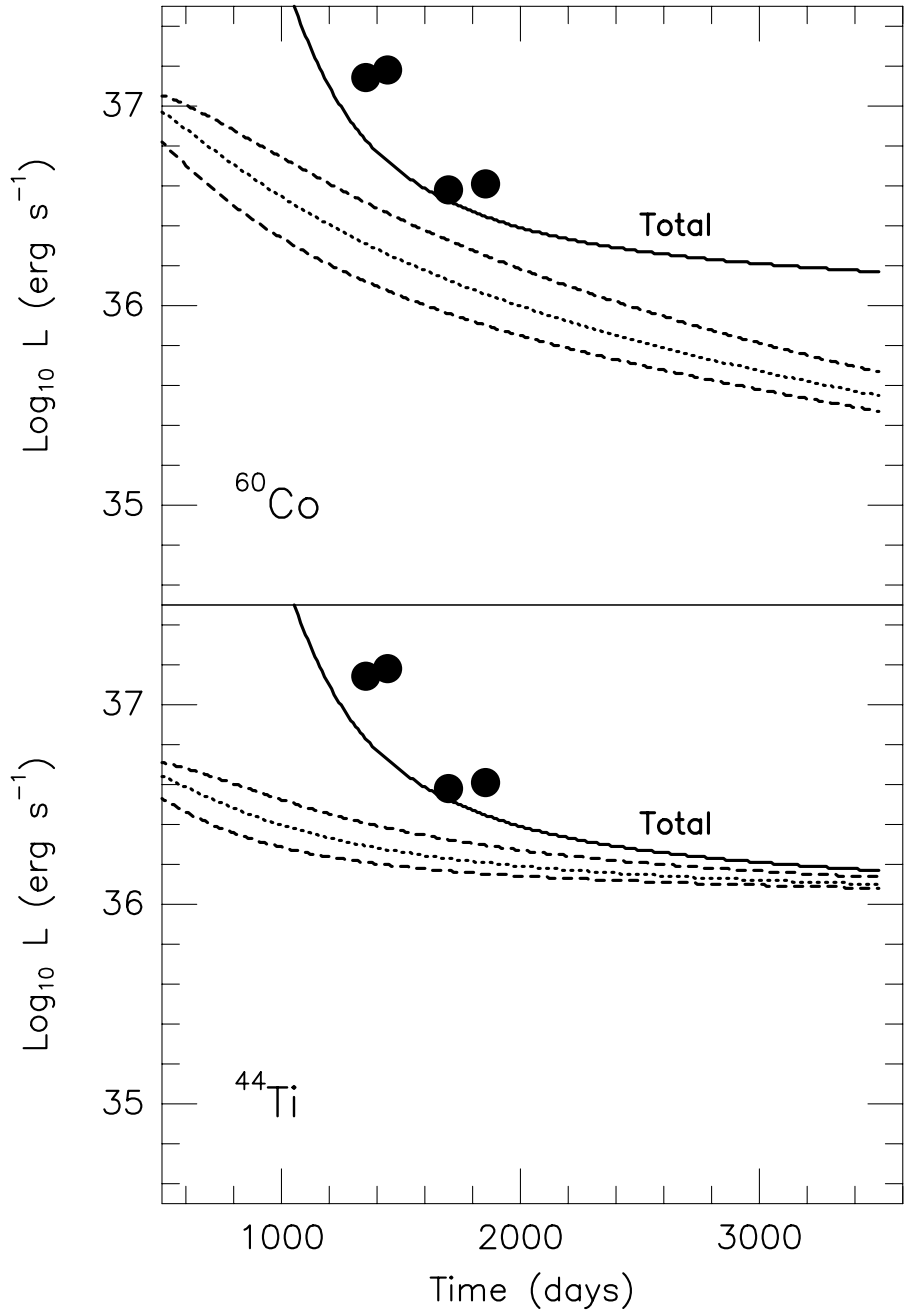

Fig. 4.— Bolometric light curve of SN 1987a from 500 to 3500 days and the effect of varying the canonical value of the effective opacity (dotted curve), $\kappa_{44}=\kappa_{60}=0.40$ cm$^2$ g$^{-1}$, by factors of two (dashed curves). The changes are largest for $^{60}$Co since it deposits a larger energy per decay than $^{44}$Ti.